\documentclass[twoside,twocolumn,english]{revtex4-2}
\usepackage[T1]{fontenc}
\usepackage[latin9]{inputenc}
\usepackage{amsmath}
\usepackage{amssymb}
\usepackage{graphicx}
\usepackage{wasysym}
\usepackage{esint}

\makeatletter
\usepackage{soul}
\usepackage{color}
\usepackage{babel}
\usepackage{calc}

\makeatother

\usepackage{babel}
\begin{document}
\title{Instability cascades in crumpling mylar sheets follow a log-Poisson statistic}
\author{Stefan Boettcher$^{1}$\email{sboettc@emory.edu} and Paula A. Gago$^{2}$\thanks{paulaalejandrayo@gmail.com}}
\affiliation{$^{1}$Department of Physics, Emory University, Atlanta, GA 30322,
USA~\\
$^{2}$Department of Earth Science and Engineering, Imperial College,
London, SW7 2BP, UK}
\begin{abstract}
The process of aging following a hard quench into a glassy state is
characterized universally, for a wide class of materials, by logarithmic
evolution of state variables and a power-law decay of two-time correlation
functions that collapse only for the ratio of those times. This stands
in stark contrast with relaxation in equilibrium materials, where
time-translational invariance holds. It is by now widely recognized
that these aging processes, which ever so slowly relax a complex disordered
material after a quench, are facilitated by activated events. Yet,
theories often cited to describe such a non-equilibrium process can
be shown to miss pertinent aspects that are inherent to many experiments.
A case in point are recent experiments on crumpling sheets of mylar
loaded by a weight whose acoustic emissions are measured while the
material buckles. Using extensive simulations to generate long time-series of 
such buckling events, we  show that crumpling
is a log-Poisson process activated by increasingly rare record-sized
fluctuations in a slowly stiffening material characterized by  a logarithmically growing length-scale. Crumpling thus adds
to a range of glassy materials exhibiting the log-Poisson property,
which can be used to discriminate between theories.
\end{abstract}
\maketitle

\section{Introduction}
\label{Introduction}
The aging process that is most concise and revealing is induced by
preparing a system at a high temperature (or low density, least stress,
etc), where the system equilibrates easily, and then instantaneously
quenching it down to a fixed, low temperature, to explore how it relaxes
towards equilibrium thereafter. Such a hard quenching protocol to
induce aging \cite{Struik78}, when applied to glass-forming systems,
elicits quite subtle relaxation behaviors which keeps the system far
from a new equilibrium for very long times. Despite vast differences
in scale and structure, there exists significant commonality in the
aging behavior of glassy systems \cite{Lundgren83,Cipelletti00,Courtland03,Buisson03,Buisson03a,Elmasri05,Cianci06,Lynch08,Yunker09,Hoffmann90,Rieger93,Crisanti04,Sibani05,Sibani06b,Christiansen08,GB20}
that is hinting at a universal origin independent of those microscopic
details, as already noted by Cottrell \cite{Cottrell52} for creep, in particular.

\begin{figure}
\hfill{}\includegraphics[viewport=0bp 240bp 792bp 570bp,clip,width=1\columnwidth]{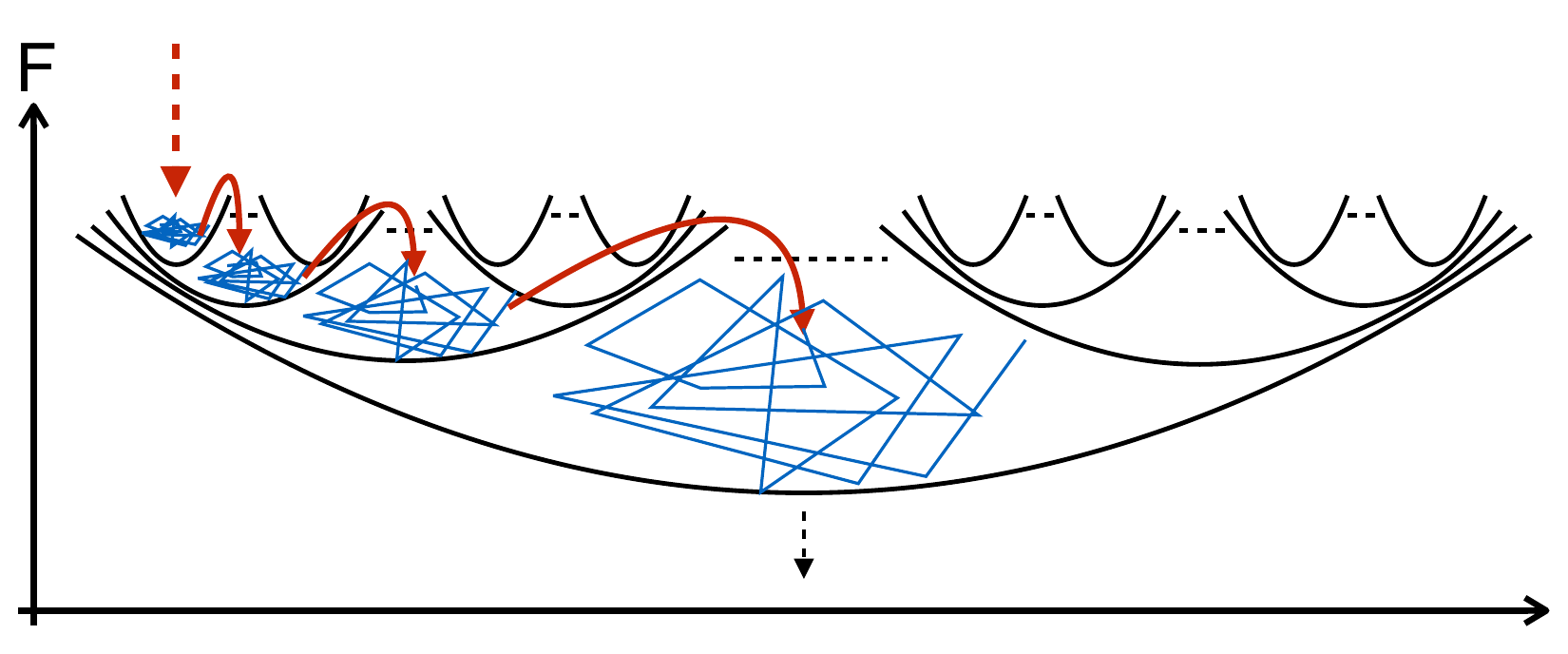}\hfill{}

\caption{\label{fig:FLandscape} Sketch of the hierarchical free-energy landscape
of a disordered system, with a typical trajectory of an aging dynamics
(blue and red) \cite{Robe16}. With increasing free energy $F$, local
minima proliferate rapidly but also become shallower. Relaxing downwards
(red-dashed arrow), the dynamics evolves through a sequence of quasi-equilibrium
explorations (blue) and intermittent, irreversible avalanches over
record barriers (red) that access an exponentially expanding portion
of configuration space (black-dashed arrow) \cite{BoSi}.}
\end{figure}

Recent experiments \cite{Shohat23,Lahini23} have probed the
aging dynamics for creep in crumpling sheets of mylar after a constant load
is applied. While prior experiments merely observed the scale of global
compression of the sheets \cite{Matan02,Lahini17}, recent work manages
to resolve the buckling of the material on a microscopic level also
by measuring the time series of acoustic emissions. As found in the
earlier work, the height of the compressing weight decreased logarithmically
with time after the start of the experiment. In addition, the time
series similarly reveals a logarithmic accumulation of clicks, at
longer times resolved into distinct cascades indicative of random
threshold events. 

Indeed, it has long been observed that the aging process in real-world
experiments is progressing via intermittent events, activated by thermal
fluctuations \cite{Courtland03,Buisson03,Buisson03a,Elmasri05,Lynch08,Yunker09,Cottrell52}.
Each event unleashes an avalanche of rearrangements that relaxes the
system. Such activation is also at the basis of the trap model, referenced in 
Ref.~\citep{Shohat23} for the crumpling experiments, that is widely used to 
describe aging through transitions at broadly distributed times in an otherwise 
memoryless renewal process \cite{Bouchaud92,Schulz14}. This theory 
lacks sensitivity to system size and misses out on the full spatiotemporal 
complexity characterizing these events that could link dynamic heterogeneity 
with intermittency. Most significantly, it fails to exhibit the log-Poisson property 
found in many aging processes \cite{Boettcher18b}. Instead, these properties can be
most generically attributed to a statistics of records generated by well-localized,
activated events \cite{Sibani03,SJ13}.
Subsequently, Ref.~\citep{Lahini23} did observe a log-Poisson statistic in the inter-event 
times $\Delta t$ of the acoustic emissions from crumpling\footnote{Note, e.g., the close 
resemblance of Figs. 1c and 2b in Ref.~\citep{Lahini23} with Fig. 5a in Ref.~\citep{Sibani18}.}. 
However, its relevance is ultimately dismissed in favor of a regular Poisson process in the 
reduced variable $\Delta t/t$ for late times $t$ after the quench. Thus, establishing the 
log-Poisson property for all time-scales would eliminate both of these theories.

In Ref.~\citep{Shohat23}, these crumpling experiments are also modeled
with molecular dynamics (MD) simulations of a $2d$ network of randomly
bistable elements. Such a model, specifically designed to model creep under 
shear, may well account for the microscopic origins of the aging phenomenology 
in a large class of systems with creep, as is recently argued for in Ref.~\citep{Korchinski25}.
However, we believe that the observed temporal and spatial complexity is even
more elementary and universal \cite{Sibani03,Becker14,Robe16}. In
fact, here we show that crumpling adds another example
of a system generically exhibiting a statistic of records that can
be analyzed as a log-Poisson process \cite{Sibani03}. Hence, it rules 
out other theories of aging, such as the trap model invoked in Ref.~\citep{Shohat23}, or
an asymptotically hyperbolic distribution of relaxation rates \cite{Amir2012}
employed in Ref.~\citep{Lahini23}. Furthermore, we demonstrate that  
this ``record dynamics'' (RD) encompasses
the description of logarithmic behavior in aging and creep attempted
in Ref.~\citep{Korchinski25}, without the need to specify the microscopic
details of the avalanche dynamics triggered by these events.

In the following, we will first recount the basic principles of RD in Sec.~\ref{RD}. Our main results, in  which we apply RD to analyze the experiments on crumpling mylar sheets, are discussed in Sec.~\ref{Crumpling Experiments}. In Sec.~\ref{ClusterModel}, we use a simple mean-field cluster model to distinguish the avalanche dynamics from the essential features of log-aging. In Sec.~\ref{Conclusions}, we conclude with a summary of our main insights and a comparison with other models of aging.

 \section{Record Dynamics}
 \label{RD}

Similar to slowly but persistently driven systems showing self-organized
criticality \cite{BTW}, which remain steadily ``on the verge of
an instability'', the concept of marginal attractor stability \cite{Tang87}
is of central importance also to aging systems. These are merely perturbed
once (i.e., sheared, quenched, loaded, etc) at time $t=0$ and left
to relax through a complex free-energy landscape, as illustrated in
Fig.\,\ref{fig:FLandscape}. Beyond any specific microscopic mechanism,
we argue that the aging phenomenology of disordered systems is unified
asymptotically by the characteristic landscape features discussed
in Fig.~\ref{fig:FLandscape}, allowing to coarse-grain trajectories
of aging systems into sequences of independent, record-sized barrier
crossings\,\cite{Robe16}. 

In equilibrium, any size of fluctuation is associated with
a proportionate motion. In an aging glass, small fluctuations remain
ineffective (``in-cage rattle'') and only large ones may amount
to irreversible (``cage-breaking'') events that actually relax the
system. Accordingly, RD associates aging with atypically large but
localized fluctuations that lead to intermittent, irreversible events
found in spin glasses, polymers, colloids, and granular piles \cite{Bissig03,Buisson03,Parker03,Sibani06a,Yunker09,Kajiya13,Zargar13,Tanaka17,GB20}
after a quench, not unlike those observed for crumpling sheets exposed
to a load \cite{Shohat23,Lahini23}. As the authors of Ref.~\citep{Shohat23}
point out, ``quiescent dwell times between avalanches grow with the
system\textquoteright s age due to a small yet steady increase in
the lowest effective energy barrier''. They report that ``after each
avalanche, the network reorganizes and slightly stiffens'', similar
to the observation in Ref.  \cite{Yunker09} for aging colloids
and reminiscent of the exhaustion theory of creep discussed in 
1952 in Ref. \cite{Cottrell52}. Ref.~\citep{Shohat23} concludes that 
``As a result, the barrier for the next event increases'',
as shown in the inset of their Fig. 3d. This implies a progression
of fluctuation records needed  to advance the relaxation dynamics, 
which is the fundamental tenet of RD \cite{Sibani03}. 

These record fluctuations drive the dynamics (i.e., ``set the clock'')
in disordered materials, generically. While this ``clock'' decelerates
as new records are ever harder to achieve, on it, dynamics proceeds
homogeneously in $\ln t$ instead of in linear time: Any sequence
of $t$ independent events produces records at a rate $\lambda(t)\sim r/t$
\cite{SJ13}. This implies for the average number of such intermittent
events (``clicks of the clock'') in an interval $\left(t_{w},t\right]$:
\begin{equation}
\left\langle n_{I}\left(t,t_{w}\right)\right\rangle =\int_{t_{w}}^{t}\lambda(\tau)\,d\tau\sim r\ln\left(\frac{t}{t_{w}}\right),\label{eq:LogPoisson}
\end{equation}
with an explicit dependence on the age $t_{w}$. Any two-time correlations
become \emph{subordinate} \cite{Sibani06a} to this clock: $C\left(t,t_{w}\right)=C\left[n_{I}\right]=C\left(t/t_{w}\right)$,
such that memory effects ensue. (This subordination corresponds to
the ``reparametrization invariance for aging systems'' noted before
\cite{PhysRevLett.89.217201}.) Compared with a Poisson statistic,
where a constant rate $\lambda$ provides time-translational invariance
(stationarity), $\left\langle n_{I}\right\rangle \propto t-t_{w}$
and an exponential distribution of inter-event times $\Delta t_{i}=t_{i+1}-t_{i}$,
RD in Eq.\ (\ref{eq:LogPoisson}) is a log-Poisson process \cite{Sibani03,BoSi09,SJ13,Boettcher18b}
with exponentially distributed values of $\Delta\ln t_{i}=\ln(t_{i+1}/t_{i})=\ln(1+\Delta t_{i}/t_{i})$.
It also shows that the rate-constant $r$ in Eq.\ (\ref{eq:LogPoisson})
is evidently proportional to the number of particles (or spins) in
view \cite{Sibani03}, as well as an increasing function of temperature
\cite{Sibani18} (or density \cite{Robe16}, etc). This $r$ ceases to exist
when the system is heated to near the glass transition, where the
observed aging behavior disappears.

\subsection{Prior observations of Record Dynamics}

While record dynamics (RD) has long been verified in simulations and
experiments on spin-glasses \cite{Sibani05,Sibani06a,Sibani07}, there
is also direct experimental evidence for RD in structural glasses
\cite{Yunker09,Kajiya13,Zargar13}. Yunker et al \cite{Yunker09}
studied rapid quenches into a high-density state of colloids in \emph{2d}
that change radius with temperature and observed the ensuing activity.
Those allow a \emph{direct} measure of record statistics \cite{Robe16},
namely, that the rate of cage-breaking events decelerates as $\approx1/t$. This is due to
the fact that later events entail minutely growing \emph{clusters}
of cooperating particles to rearrange \cite{Yunker09,Tanaka17}, as
shown in Fig.~2b of Ref.~\citep{Yunker09}, to scratch together
enough free volume to move within an increasingly stiffened (aged)
environment. A similar deceleration of events was also observed by
Kajiya et al \cite{Kajiya13}. Such cage breakings in colloidal suspensions
are irreversible, pace-setting events associated with crossings of
free-energy barriers \cite{Hunter12}. 

Similar intermittent events are common to other aging glassy materials
such as quenched spin glasses \cite{Sibani05} and compactifying granular
piles \cite{GB20}. There, tapping frees up shrinking bits of free
volume over expanding domains at a $\sim1/t$ rate so that the accumulated
expelled volume (i.e., the loss of free energy) increases as $\sim\ln t$
and, in turn, saturates the density ever-so-slowly, $\sim1/\ln t$
\cite{Nowak98,Zargar13,Sibani16,GB20}, similar to earlier Lennard-Jones
simulations by El-Masri et al \cite{ElMasri10}, or cluster model
predictions \cite{Becker14}.
In spin glasses, increasingly rare barrier crossings require minutely
growing clusters of spin-flip events \cite{Sibani18}. 

A direct observation of the log-Poisson statistics that is implied
by RD often requires more extensive sampling and/or spatial resolution than
many experiments can provide. However, large-scale simulations of colloids
and spin glasses, for instance,
where those specific localized events can be easily resolved~\citep{Boettcher18b}, have
provided ample evidence for the log-Poisson statistic as well as its
scaling collapse with domain size, similar to Fig.~\ref{fig:logPoissonCrumble}  for crumbling.

The proportionality of the event statistics to system size $N$ provides
a very sensitive test for a log-Poisson process, as it should
allow to collapse data measured for the time series of events at different
sizes when rescaled by $N$, as shown in Ref.~\citep{Boettcher18b}
and demonstrated for the bistable model of crumpling below. Such a
dependence, inherent in RD, is absent from other theories that assume
the thermodynamic limit and thereby average out any notion of a localized
event.

\begin{figure}
\hfill{}\includegraphics[viewport=0bp 20bp 735bp 550bp,clip,width=1\columnwidth]{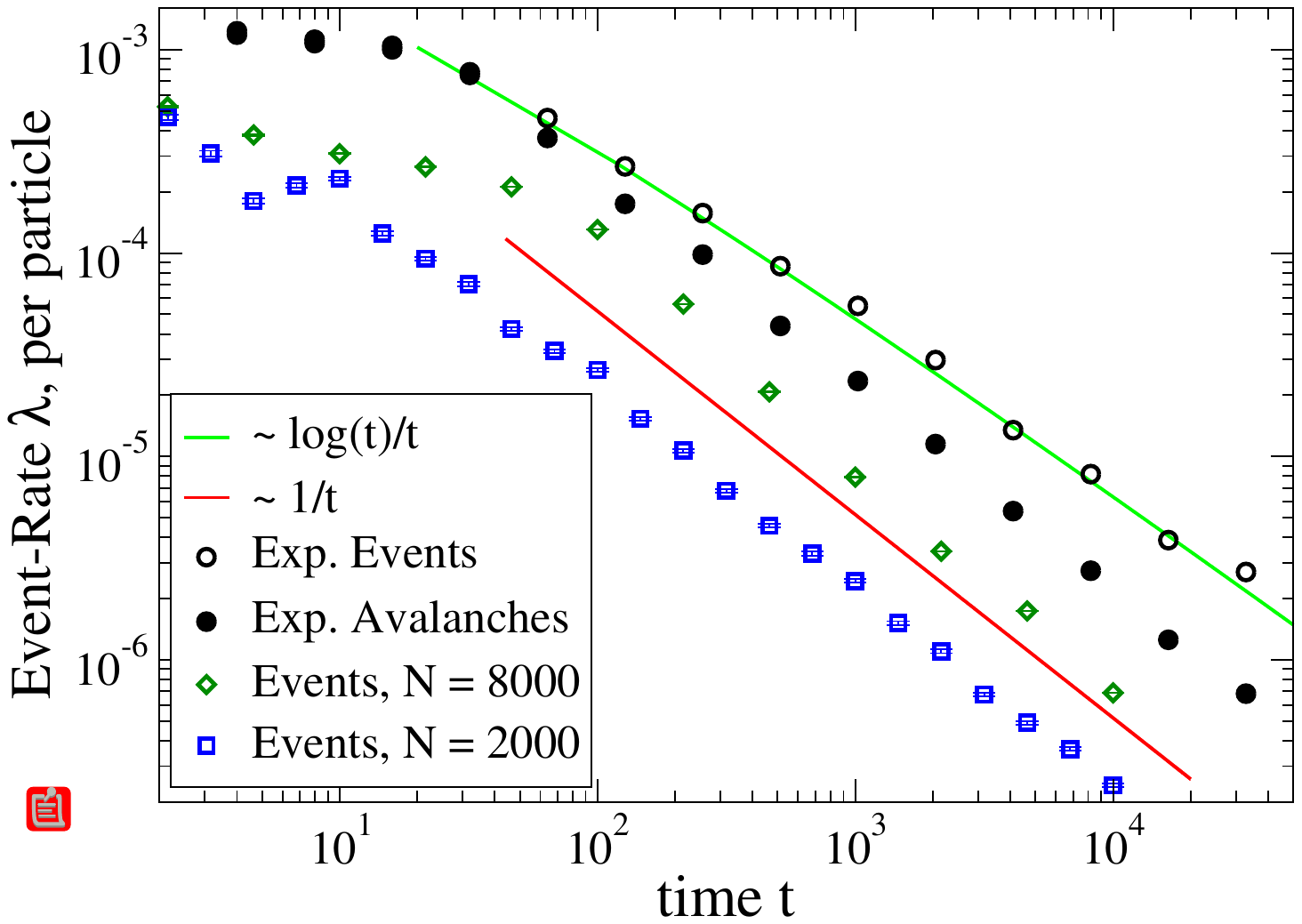}
\hfill{}

\vspace{-0.5cm}

\caption{\label{fig:Decay}{\small{} Decay in the rate $\lambda(t)$ of intermittent
events (``clicks'') after applying a load to crumpled sheets in
the experiments (circles, in arbitrary units) and for the MD simulations
(squares for $N=2000$ and diamonds for $N=8000$ nodes, averaged
over 2000 instances) of the bistable model, both provided in Ref.~\citep{Shohat23}.
Counting all events in the experiments ($\Circle$) appears to result
in a logarithmic overcount on record statistic, i.e. $\lambda(t)\sim\log(t)/t$
(green line), which is removed by combining all subsequent events
within a time window of $\delta t/t=0.001$ after the onset into a
single avalanche ($\CIRCLE$) that occur at rate of $\lambda(t)\sim1/t$
(red line). In the simulations, the rate for all events are consistent
with the hyperbolic behavior predicted by RD, even without corrections.}}
\end{figure}

We already have found direct evidence for RD and log-Poisson statistic
in various disordered materials such as in experiments and simulations
on spin glasses \cite{Sibani06a,Sibani18} and on colloids \cite{Yunker09,Robe16},
as well as in simulations of granular piles \cite{GB20}. In the following, we will detail the evidence for RD
in the crumpling experiments.

\section{Crumpling of Mylar Sheets}
\label{Crumpling Experiments}

With the experimental data and the MD
implementation of the bistable model that the authors of Ref.~\citep{Shohat23}
have generously provided in their supplemental material, we also reproduced
the main features of RD for crumpling. In particular, we directly measured the 
hyperbolic deceleration of the experimental event rates, as discussed in 
Fig. \ref{fig:Decay}. With the bistable model, we obtain the log-Poisson
distributions in the variable $\Delta\ln t_{i}$ up to 
$\ln\left(1+\Delta t_{i}/t_{i}\right)\approx6$, i.e., $\Delta t_{i}/t_{i}\approx400$, 
and their collapse with system size, as shown in Fig. \ref{fig:logPoissonCrumble}. 
(Details of our simulations are provided in the Appendix.)
Note that this result is inconsistent with the trap
model cited in Ref.~\citep{Shohat23} as well as with the alternative
prediction given in Ref.~\citep{Lahini23} of a Poissonian exponential
distribution for $\Delta t_{i}/t_{i}$ derived from the theory in
Ref.~\citep{Amir2012}, which merely coincides with the actual log-Poisson
statistic for $\Delta t_{i}/t_{i}\ll1$.

\begin{figure}
\hfill{}\includegraphics[viewport=0bp 200bp 450bp 510bp,clip,width=1\columnwidth]{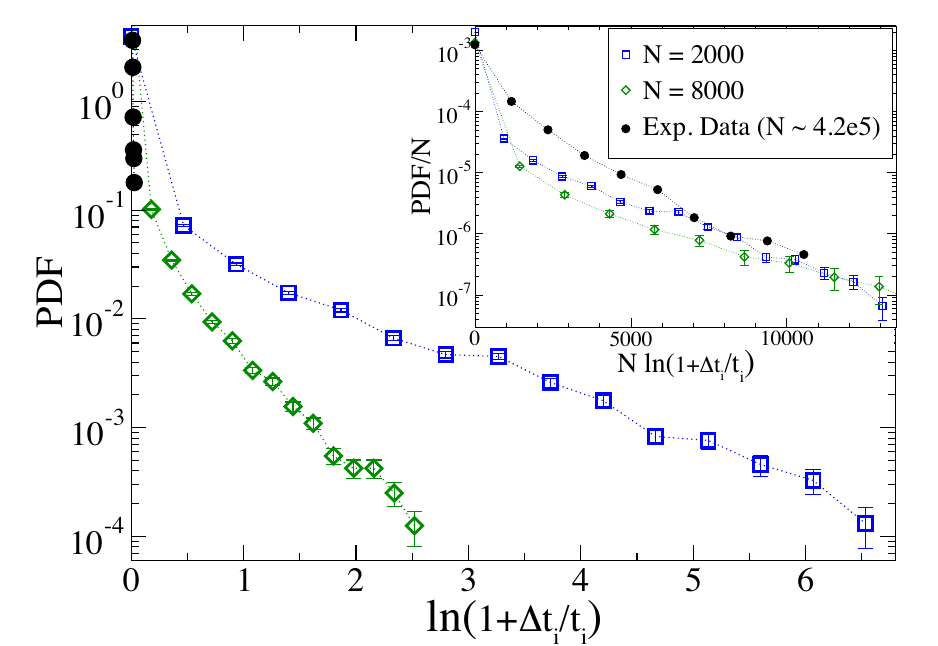}
\hfill{}

\vspace{-0.5cm}

\caption{\label{fig:logPoissonCrumble} Log-Poisson statistic for the inter-event
intervals $\Delta\ln t_{i}=\ln\left(1+\Delta t_{i}/t_{i}\right)$
for the $i^{{\rm th}}$ event, measured within two different system
sizes, $N=2000$ and $N=8000$, for the MD simulations as well as
the experimental data provided by Ref.~\citep{Shohat23}. The main
plot shows the raw data for each PDF, while the collapses of that
data is shown in the inset, rescaled by the system size $N$. Accordingly,
the experimental data would correspond to a simulation with $N_{{\rm exp}}\approx4.2\ 10^{5}$
nodes. Note that the data for the small systems extends to well above
$\Delta t_{i}/t_{i}=1$, while for the experimental data it is limited
to $\Delta t_{i}/t_{i}\ll1$ \cite{Lahini23}. The simulation data
in Ref.~\citep{Shohat23} is plotted for $\Delta t_{i}/t_{i+1}$,
which is strictly $\protect\leq1$.}
\end{figure}

\begin{figure}
\hfill{}\includegraphics[viewport=0bp 100bp 580bp 510bp,clip,width=1\columnwidth]{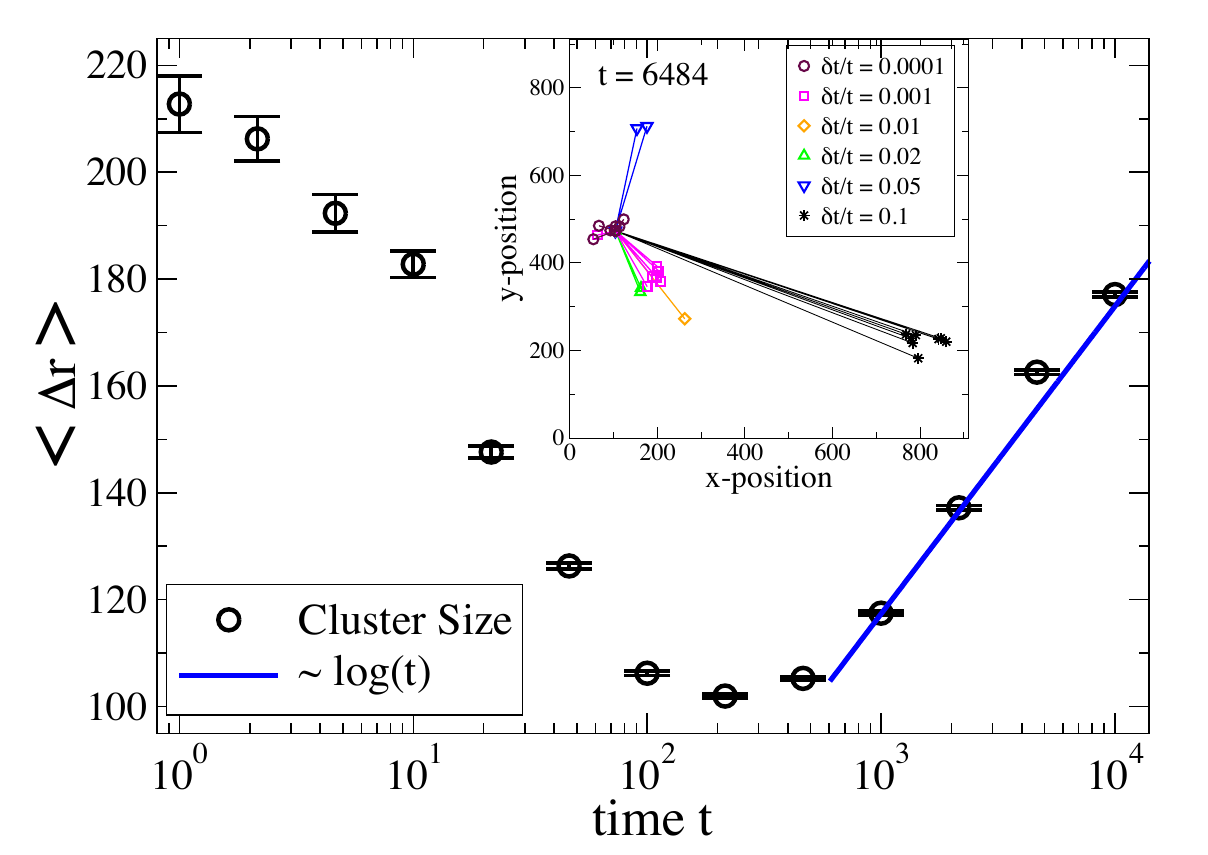}
\hfill{}

\vspace{-0.5cm}

\caption{\label{fig:clustersize}{Average spatial extend $\left\langle \Delta r\right\rangle $
of the clusters of events within a time window $\delta t$ following
an initial event at time $t$ in MD simulations of the bistable 
model with $N=8000$ nodes within a 900-by-900 box (see inset), here
for $\delta t/t=0.001$. For short times immediately after applying
the load ($t<200$), too many otherwise independent events that are
spread randomly over the box overlap to artificially enlarge $\left\langle \Delta r\right\rangle $.
With decelerating activity, these overlaps fade to reveal the logarithmic
growth of authentically correlated behavior within those windows,
as predicted for RD \cite{Becker14,Robe16,GB20}. The inset illustrates
such a spatially correlated cluster for a typical (albeit unusually
active) sequence of events, initiated at a late time $t=6484$ at
coordinate $\left(104,473\right)$. Depending on window size $\delta t$
(see legend), an increasing set of additional events -- increasingly
farther apart -- are taken as correlated with the first event, suggesting
reasonable intermediate choices for $\delta t$ that provide robust
log-scaling (not shown).}}
\end{figure}

Although it runs somewhat counter to the cherished equilibrium notion
of the thermodynamic limit, to detect the log-Poisson property it
is essential to find a system (or independent domains therein) small
enough, or aged enough, to sufficiently separate avalanche
events. For a system far from equilibrium that evolves intermittently,
overlapping too many events from far-flung, independent domains corresponds
to superimposing multiple time series and thereby curtailing the possibility
to observe events with larger values of $\Delta t/t$. We imagine
it to be difficult to separate events from different domains within
the actual crumpling experiments, short of repeating the crumpling
with smaller sheets or loads, say. However, comparing simulations of systems with
$N=2000$ and $8000$ nodes in Fig. \ref{fig:logPoissonCrumble} shows
not only the log-Poisson property of the inter-event statistic but
also an asymptotic collapse when we rescale simply by $N$ (see
inset), as would be expected for combining time series from simultaneously
but independently evolving log-Poisson processes \cite{Boettcher18b}.
Yet, blocking the entire system of $N=8000$ nodes into squares, each
containing $N/4=2000$ nodes, did not provide a good collapse, unlike
for other systems \cite{Boettcher18b}, hinting at longer-range correlations
via force chains \cite{Daniels17} that arise quite naturally in an
elastic system. Such correlations in the aging behavior are owed to
the driving force exerted on the walls that pervade the entire system,
unlike the purely local, thermally-activated aging in a quenched system.
That this range becomes rather large in this model is also evident
from Fig. \ref{fig:clustersize}, although the largest range observed
($\approx180$ units at the largest time) is still small compared
to the box width, $L=900$, containing the system.

Clearly, not every acoustic emissions is an event in the sense of
RD. Rather, isolated sequences of causally related emissions can be
grouped into avalanches, which \emph{collectively} represent irreversible
record events. Often, these spurious events, internal to each avalanche,
can be ignored, since their inter-event times are so small such that
$\Delta t_{i}/t_{i}\approx0$, merely contributing to a peak near
$\Delta\ln t_{i}=0$, as can be seen also in Fig.\ref{fig:logPoissonCrumble}.
However, this overcount can at times skew the statistics because it
affects the norm of the log-Poisson PDF or it leads to log-corrections
to the event-rate, see Fig. \ref{fig:Decay}. Thus, we introduce some
time window $\delta t\propto t$, which is proportional but small
relative to $t$ after an initial event initiated at time $t$, to
suppress immediately subsequent events that we deem as causally related.
The actual width of this window is due to unknown microscopic details of the 
avalanche process. However, as long as we chose $\delta t/t$ sufficiently large 
to eliminate the distortions from the $1/t$-scaling, but small enough to resolve 
most avalanches and obtain enough statistics, the specific choice has no effect 
on the long-time behavior (see Fig. \ref{fig:clustersize}). In Fig. \ref{fig:Decay},
we find that such a window is useful for the experimental data to
discover the true RD by removing a logarithmic overcount of events,
while in the simulations there is hardly any measurable effect comparable in size
to those experimental ``aftershocks''. (These aftershocks are reminiscent
of Omori's law for earthquakes, whose rate also decays hyperbolically
in $t$ \cite{Helmstetter02}.) 

\begin{figure*}
\hfill{}\includegraphics[viewport=0bp 320bp 800bp 530bp,clip,width=1\textwidth]{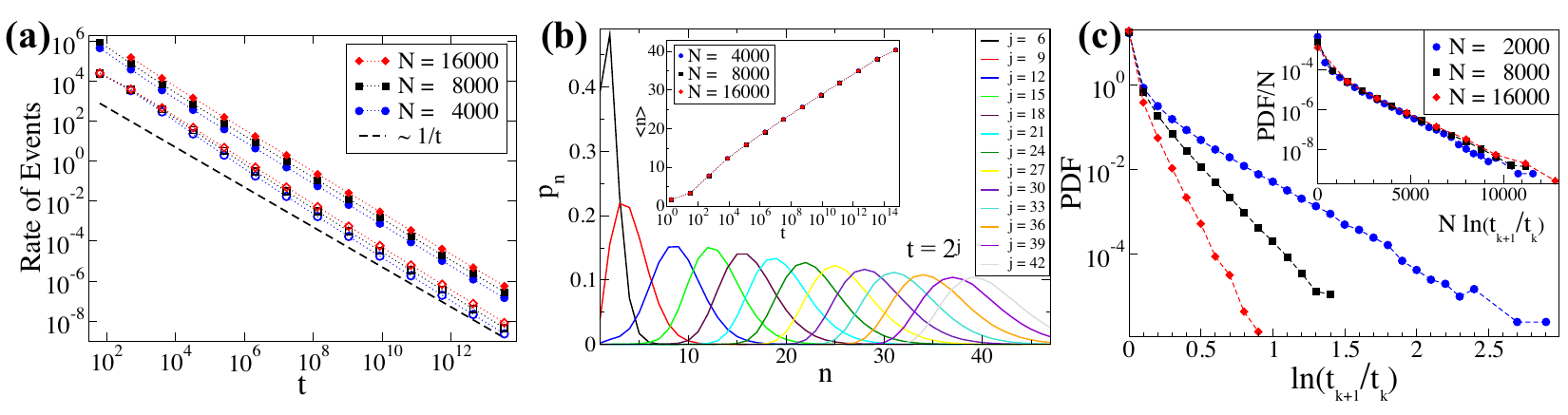}\hfill{}

\vspace{-0.5cm}

\caption{\label{fig:OnlyClusterPlots}Generic results for the mean-field cluster
model \cite{BoSi09,Becker14} of size $N$, here obtained by randomly
partitioning clusters of size $n_{k}>1$ into two. (a) Hyperbolically
decelerating rate of break-up events, whether all events are counted
(solid symbols) or discounting those within a window of $\delta t=0.01t$
following an event at $t_{i}=t$ (open symbols). (b) Distribution
$p_{n}$ of clusters of size $n$, measured at an exponential progression
of times $t=2^{j}$ for $j=6,9,\ldots,42$ at $N=8000$, that shows
the log-time expansion of a ``gap'' below the smallest of the large
clusters at any $t\gg1$, such that the average cluster size grows
as $\left\langle n\right\rangle \sim\ln t$ independent of $N$ (see
inset), similar to Fig. \ref{fig:clustersize}. (c) Log-Poisson statistic,
and its collapse when rescaled with $N$ (inset), comparable with
the crumpling simulations in Fig. \ref{fig:logPoissonCrumble}. Note
that the overcount of many events with small $\Delta t$ inside avalanches
at large $t$ merely result in a spurious peak at the origin.}
\end{figure*}

\begin{figure}[b!]
%\vspace{-0.5cm}
\hfill{}\includegraphics[viewport=0bp 10bp 740bp 540bp,clip,width=0.43\columnwidth]{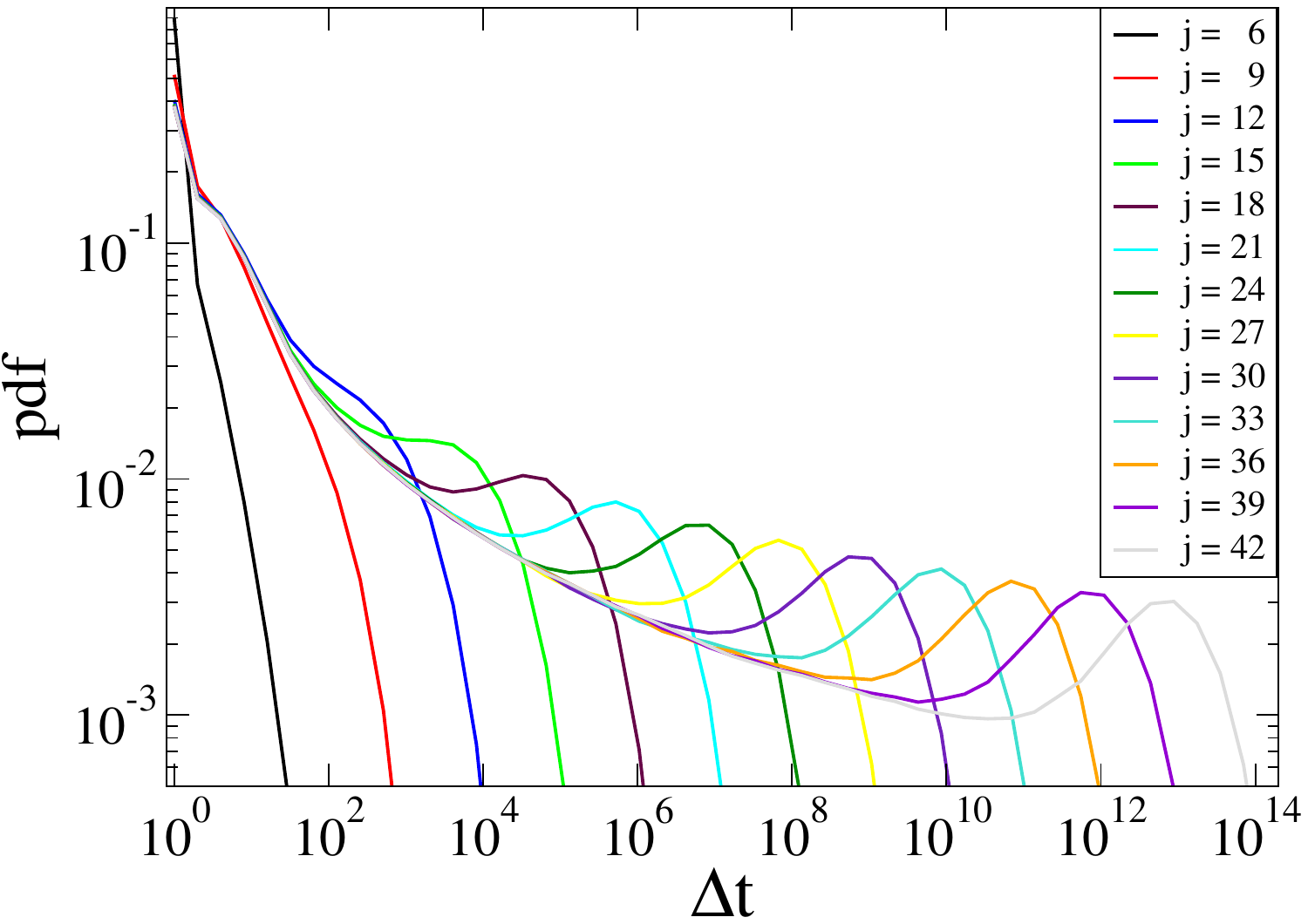}\hfill{}\includegraphics[viewport=0bp 170bp 620bp 520bp,clip,width=0.55\columnwidth]{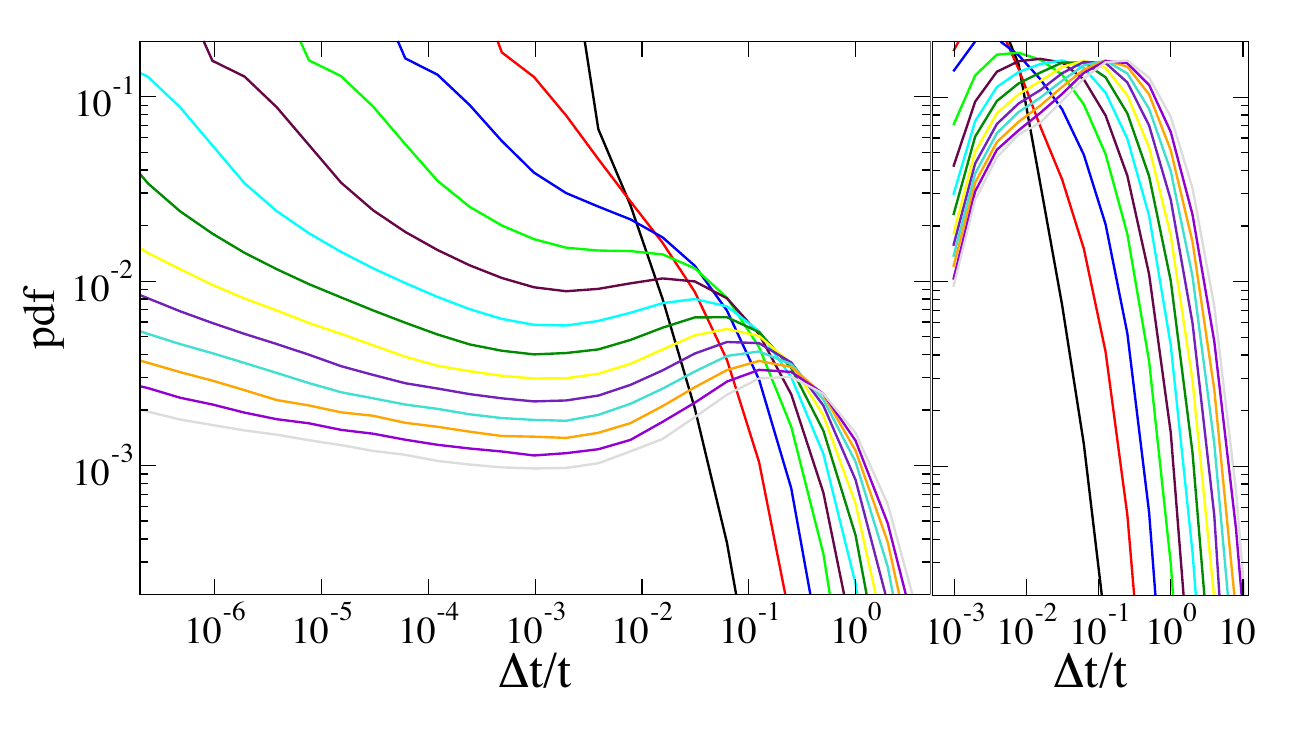}\hfill{}

\hfill{}\includegraphics[viewport=0bp 10bp 740bp 540bp,clip,width=0.43\columnwidth]{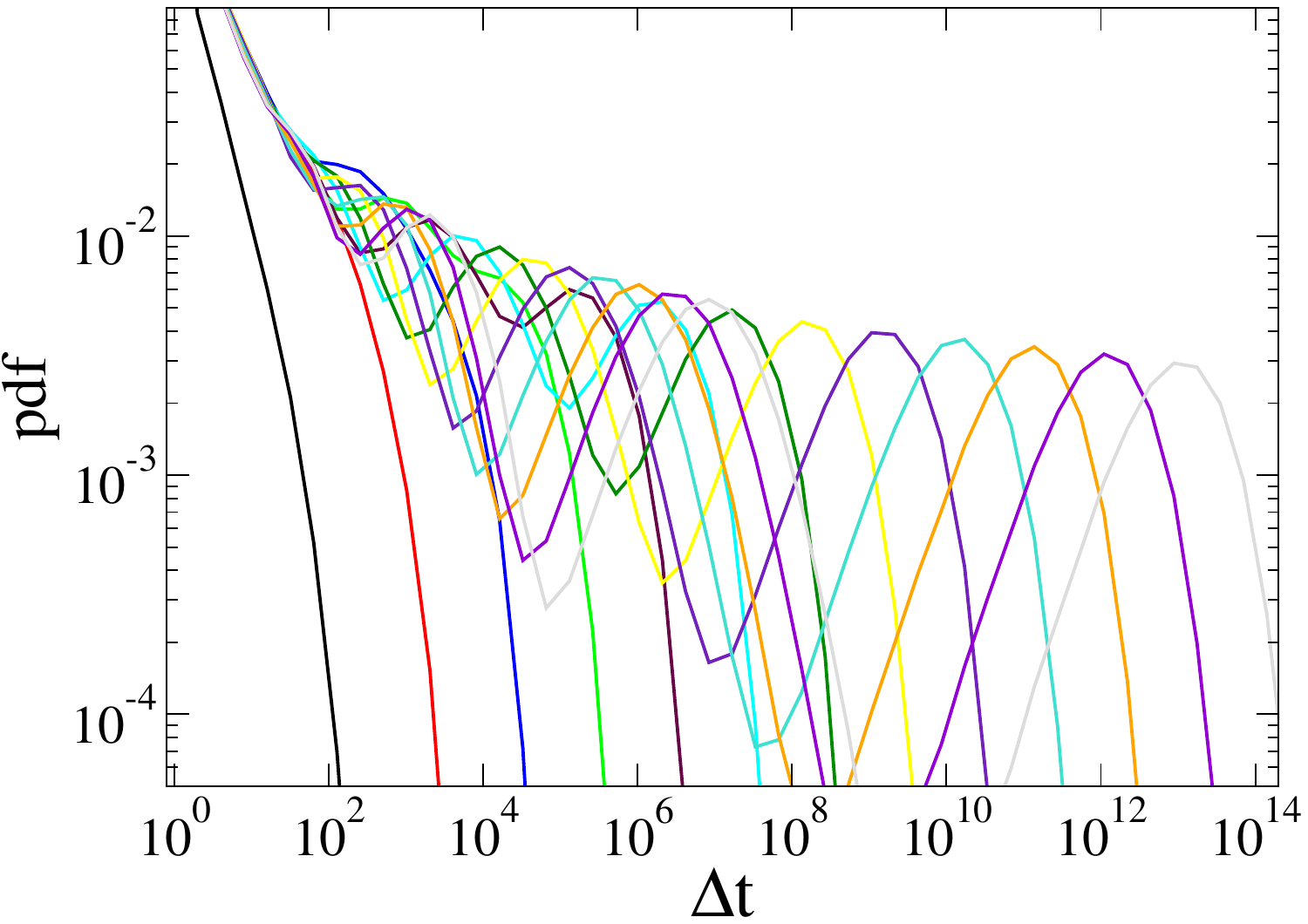}\hfill{}\includegraphics[viewport=0bp 220bp 545bp 520bp,clip,width=0.55\columnwidth]{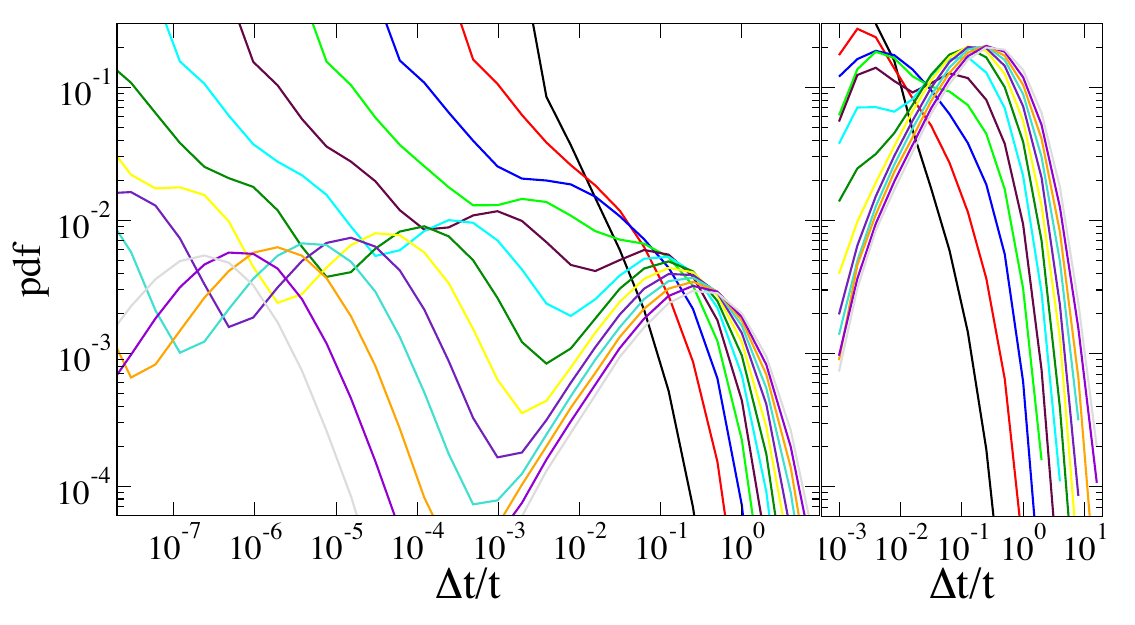}\hfill{}

\hfill{}\includegraphics[viewport=0bp 10bp 740bp 540bp,clip,width=0.43\columnwidth]{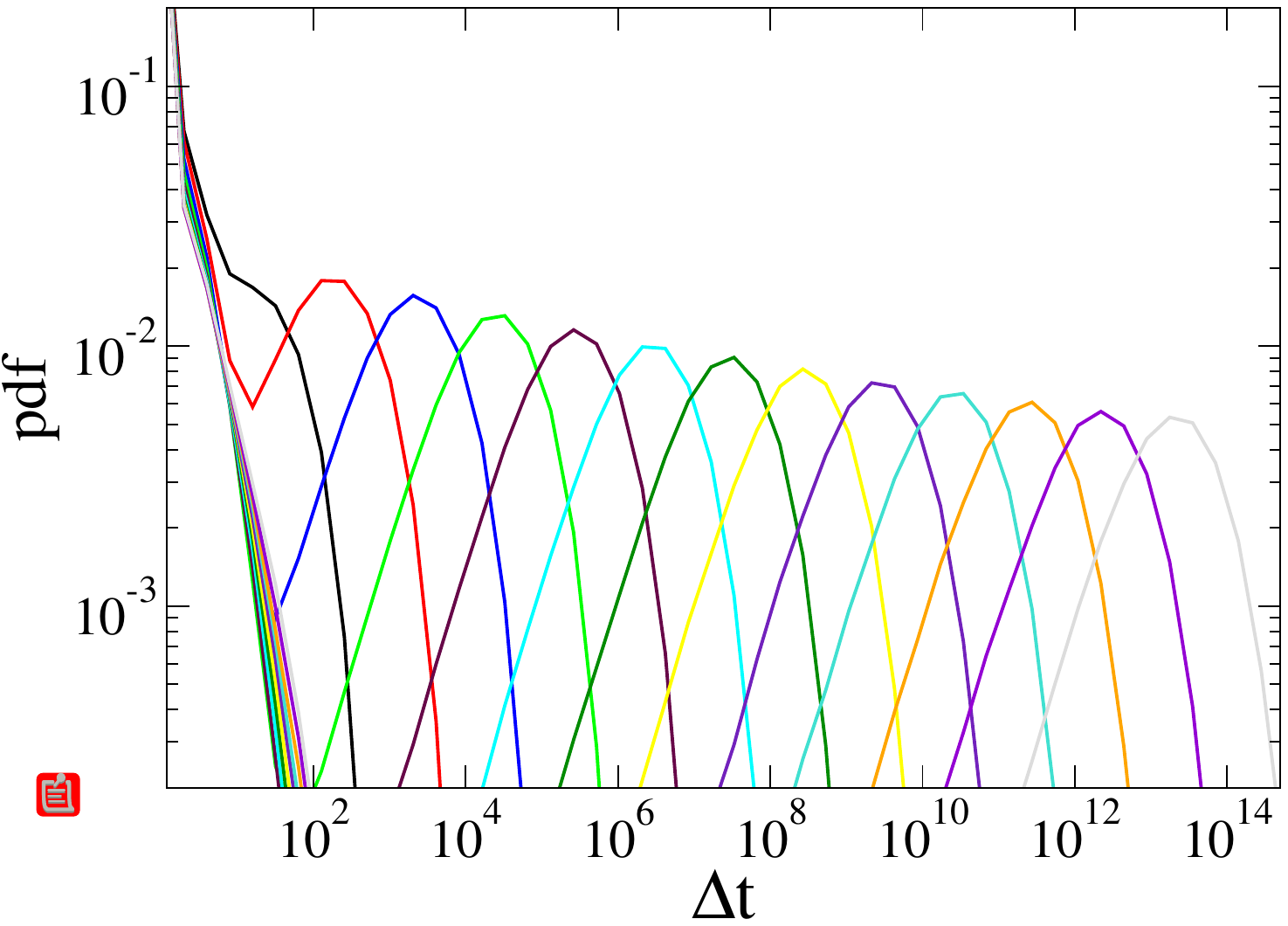}\hfill{}\includegraphics[viewport=0bp 220bp 540bp 520bp,clip,width=0.55\columnwidth]{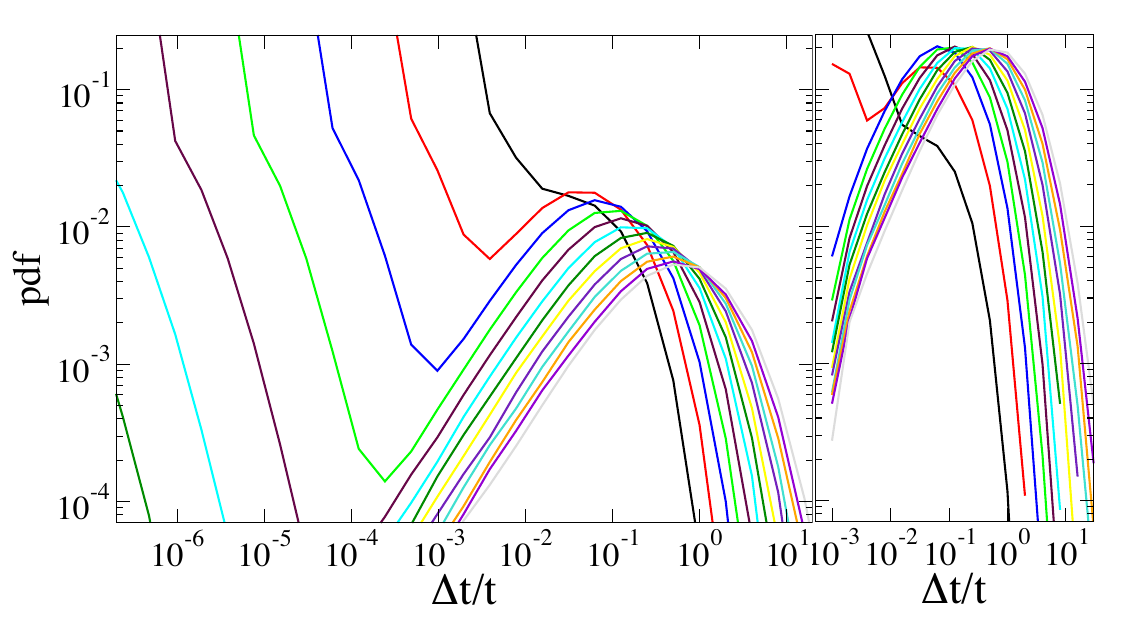}\hfill{}

%\vspace{-0.3cm}
\caption{\label{fig:IET}Probability density functions $P_{t}(\Delta t)$ and
$P_{t}(\Delta t/t)$ for the inter-event times $\Delta t$ following
events occurring at a system age $t$ with $2^{j}<t\protect\leq2^{j+3}$
for three different break-up methods in the mean-field cluster model.
In all cases, the system size is $N=8000$. Data in the top row refers
to a random bi-partitioning of the cluster that is breaking up. For
the middle row, each break-up leads to two equal-sized clusters (or
$\pm1$ in relative size). For the bottom row, each break-up of a
cluster of size $n$ immediately results in $n$ new clusters of unit
size. The leftmost panel for each method shows $P_{t}(\Delta t)$,
which is plotted rescaled as $P_{t}(\Delta t/t)$ in the right panels,
where the rightmost panel shows the corresponding data for the case
where all subsequent dependent events within a time-window of $\delta t=0.001t$
after the (presumably originating) event at $t$ are suppressed. (The same legend
applies to all plots.)}
\end{figure}

\section{A Cluster Model Analysis of the Avalanche Dynamics}
\label{ClusterModel}

Ref.~\citep{Korchinski25} attempts
to relate the internal dynamics of these avalanches to the aging phenomenology.
In RD, these avalanches are just coarse-grained into instantaneous
events, as it is sufficient for logarithmic aging that such a sequence
of ever larger events exists, irrespective of their internal dynamics.
To demonstrate that the internal dynamics of the avalanches is not
relevant for the aging phenomenology, we discuss a simple model of
a record process. (This is a mean-field version of the cluster model
introduced in Ref.~\citep{BoSi09} and summarized in the Appendix.) $N$ particles are partitioned into clusters, 
whose size is signifying
the increasingly stiffening domains in the aging system, that have
variable sizes $n_{k}$ ($\sum_{k}n_{k}=N$) with Poissonian distributed
lifetimes $\tau_{k}$ of average $\tau_{{\rm m}}e^{\beta n_{k}}$.
(Here, we simply set $\tau_{{\rm m}}=1$ for some microscopic time-scale
and $\beta=1$.) The system is initiated at $t=0$ (the ``quench'')
with all particles being isolated, i.e., $n_{k}=1$ for all $k$.
At each update $i$, we select that cluster $\kappa$ with the shortest lifetime,
$\Delta t_{i}=\min_{k}\left\{ \tau_{k}\right\}=\tau_\kappa $, advance the system
time to $t=t_{i}+\Delta t_{i}=t_{i+1}$ and the lifetimes for each
cluster to $\tau_{k}\to\tau_{k}-\Delta t_{i}$. Then, we execute one of
these two procedures: (1) If cluster $\kappa$ is an isolated particle
($n_{\kappa}=1$), then this particle joins a random cluster 
$k^{\prime}\not=\kappa$, i.e., $n_{k^{\prime}}\to n_{k^{\prime}}+1$ and the total
number of clusters decrements, or otherwise (2) cluster $\kappa$ breaks
up in some form, momentarily increasing the number of clusters. This break-up
(the ``avalanche'') may lead to two new clusters by randomly partitioning
$n_{\kappa}$, say, or to $n_{\kappa}$ new unit-sized clusters, etc; the very 
point is that these avalanche details don't matter 
for the aging behavior. After each update, any newly created
clusters receive a new lifetime and the next update ensues. 

\subsection{Generality of the emerging Gap in Record Dynamics}
\label{gap}
Crucially, small clusters -- and especially isolated particles --
receive exponentially smaller lifetimes than larger ones and get ``mopped
up'' rather quickly. Thus, most events are of type (1), yet, even
a cascade of these small events happens almost instantaneously on 
the scale of the system time $t$, i.e., $\Delta t_{i}\ll t_{i}$, and adds little
to advancing $t$. What remains are the record-sized events of scale 
$\Delta t_{i}\sim t_{i}$ that dominant the tail of the inter-event
times. These concern typically the smallest of the remaining large
clusters, thus, driving the distribution of clusters minutely to ever
fewer numbers with ever larger sizes \cite{BoSi09}, which is the
evolving ``gap'' that takes a central role in Ref.~\citep{Korchinski25}.
As such a break-up typically requires a larger waiting time than any
previous, we have indeed generated a record dynamics, as Fig. \ref{fig:OnlyClusterPlots}
demonstrates. The rate of such break-up events decelerates as $\sim1/t$
asymptotically, whether the avalanche events are counted or not (Fig.
\ref{fig:OnlyClusterPlots}a). The cluster-size distribution develops
a gap and travels like the ``wave'' predicted in Ref.~\citep{BoSi09}, 
with log-time speed towards larger sizes (Fig. \ref{fig:OnlyClusterPlots}b)
such that the average cluster size increases as $\left\langle n\right\rangle \sim\ln t$
(inset). Finally, inter-event
times $\Delta t_{i}$ follow a log-Poisson statistic $\Delta\ln t_{i}$
aside from some noise near $\Delta\ln t_{i}=0$ (Fig. \ref{fig:OnlyClusterPlots}c)
which shows data collapse when rescaled by system size $N$ (inset
of Fig. \ref{fig:OnlyClusterPlots}c).

\subsection{Distribution of Inter-Event Times}
\label{IET}
Here, we discuss
the properties of the waiting-time distribution $P(\Delta t_{i})$, which probes
the dynamics internal to the avalanches and depends on the specific
break-up mechanism. Ref.~\citep{Korchinski25} emphasizes the importance 
of $P(\Delta t_{i})$ for the log-time aging phenomenology, 
which they claim to be of a universal, power-law form. In contrast, we do not see any 
relevance in its properties, aside from its cut-off scaling with system age $t$. In fact, 
even our simple examples show already that neither the avalanche dynamics nor the 
waiting-time distribution have to be critical (i.e., broadly distributed) or universal 
for log-aging to emerge. The true unifying feature is the log-Poisson statistic.

Specifically, we have sampled the properties
of the distribution $P_{t}(\Delta t)$ of inter-event times $\Delta t$
following events occurring at a system age $t$ with $2^{j}<t\leq2^{j+3}$, as 
shown in Fig. \ref{fig:IET}. We have considered different modes
by which clusters break up, each leading to a microscopically distinct
avalanche process by which the barrier crossing event gets dissipated,
which governs the core of the distribution. However, despite those
distinctions in their form, each set of distributions exhibits the
two features that we perceive as the only ones relevant for the logarithmic
aging behavior. First, for any method, the distributions for $P_{t}(\Delta t)$
exhibit a cut-off at $\sim t$ such that the cut-offs align when the
data is plotted with the reduced time variable, $P_{t}(\Delta t/t)$.
The collapse is not perfect but cut-offs are getting progressively
closer for larger $t$, demonstrating the effect asymptotically. Secondly,
this collapse in $\Delta t/t$ is unaffected when we remove all
events within a window of $\delta t=0.001t$ following an initial
event at system age $t$, supposing that those events are merely causally
dependent events within the avalanche kicked off at $t$. In fact,
although each method leads to quite distinct features for the probability
density functions at values of $\Delta t\ll t$, for $\Delta t\approx t$,
all distributions behave about the same and remain so even when
the avalanche-internal events are removed. Accordingly, all methods
lead to about the same phenomenology described for the cluster model
above. Note that none of the break-up processes (i.e.,
the avalanche dynamics) were in any way broadly distributed, as assumed 
in Ref.~\citep{Korchinski25}.

\section{Conclusions}
\label{Conclusions}

In conclusion, we have shown that the aging process in crumpling sheets
exhibits the log-Poisson property and that the data collapse when rescaled
with system size. It makes the process most consistent with a model
based on a statistic of records. While the intermittency is ``reminiscent
of the phenomenology of trap models for aging'', as Ref.~\cite{Shohat23} suggests,
that model must be rejected because it lacks the log-Poisson property
\cite{Boettcher18b}. Similarly, the model provided in Ref.~\citep{Lahini23},
is also inconsistent with this property,
as the authors themselves point out. A description of crumpling
as a record dynamics is also advantaged over that in Ref.~\citep{Korchinski25},
since it does not rely on  a microscopic analysis of the avalanche dynamics, merely
its existence, to obtain logarithmic aging.

\paragraph*{Acknowledgements:}

SB thanks the authors of Ref.~\citep{Korchinski25} for enlightening
discussions. We are grateful to the authors of Ref.~\citep{Shohat23}
for providing a rich set of data and model scripts in their supplement,
and for their detailed responses to our inquests. MD simulations of the bistable
model  were performed at the Imperial College Research Computing Service
(see DOI: 10.14469/hpc/2232). Most of the ideas
developed in this Letter have originated with our collaborator and
dear friend, the late Paolo Sibani. 

\section*{Appendix A: Cluster Model}

As a real-space incarnation of RD, a simple lattice model was designed
\cite{BoSi09,Becker14} that captures the combined temporal \emph{and}
spatial heterogeneity found in a colloidal system. The cluster model
discussed in Sec.~\ref{ClusterModel} is essentially a mean-field version of
this lattice model. Mobile particles accrete into jammed clusters
only to be re-mobilized in a chance fluctuation (``quake'') after
a time \emph{exponential} in the size of the cluster. Following a
quench at $t=0$, when all particles are mobile, clusters form and
break up \emph{irreversibly} and distribute their particles to adjacent
clusters; their growth, in turn, decelerates the dynamics. The algorithm
is exceedingly simple and consists of only \emph{two} choices \cite{Becker14},
yet, \emph{all} known experimental and MD-data for such quenches is
reproduced \cite{Robe16}. Particles always completely fill a lattice,
one on each site, $N=L^{d}$, but each particle either (1) is mobile
($h=1$), or (2) it is jammed in a cluster of size $h>1$ with adjacent
particles. When picked for an update, 
\begin{enumerate}
\item ($h=1$): a \textbf{mobile particle swaps position} with a random
neighbor and joins the neighbor's cluster; or
\item ($h>1$): a \textbf{locked-in particle creates a quake} with a cluster-size
dependent probability per sweep \cite{BoSi09},
\end{enumerate}
\begin{equation}
P_{\alpha}(h)\propto\left(1+\frac{\beta h}{\alpha}\right)^{-\alpha}\quad\xrightarrow{\alpha\to\infty}\quad P_{\infty}(h)\propto e^{-\beta h}.\label{eq:Ph}
\end{equation}
Here, $\alpha$ merely provides a convenient but unphysical means
to approach a glassy system needed below. We have also added a parameter
$\beta>0$ to eventually control the degree of jamming in the system,
similar to the area fraction (or temperature) attained in a quench.
Within clusters, fast events (cage-rattlings, spin-flips, etc), perceived
as leading to the re-arrangements preceding a quake, are intentionally
coarse-grained out, with their collective effect replaced by $P_{\alpha}(h)$.
(Such non-local effective probabilities are commonplace, e.g, in studies
of fragmentation \cite{Redner88,Krapivsky94} and coalescence \cite{Krapivsky91}.)
In this phenomenological way, our model can access behavior found
in both, structural as well as quenched glasses. 

For $\alpha=1$ or small $\beta$ in Eq.\ (\ref{eq:Ph}), a time-homogeneous
dynamics (``diffusion'') ensues with small transient clusters, at
best. Mean-square displacement (MSD) is linear in $t$, and a stationary
Poisson process results. The cluster model behaves like a liquid.
In turn, $\alpha\to\infty$ (and $\beta$ not too small) represents
a jammed colloid with full aging. Long-lived clusters emerge, whose
collective activation becomes \emph{exponentially unlikely} with their
size $h$ \cite{BenNaim98,Sibani16}. After the (record-sized) break-up
of a cluster at a time $t_{{\rm w}}$, debris almost instantly (relative
to $t_{{\rm w}}$) re-attaches onto others in an avalanche of small
events, leaving one less cluster with only a \emph{minute} increase
in $\left\langle h\right\rangle$ for those that remain. 

Marginal stability \cite{Tang87} is ensured because the next record then almost
certainly exceeds the previous only minutely and a non-trivial cluster-size
distribution with a logarithmically growing gap emerges in time \cite{BoSi09}.
These events indeed decelerate as $\sim1/t_{{\rm }}$, as predicted
by RD, which has been directly verified in colloidal experiments \cite{Yunker09,Robe16}
and our recent MD-simulations \cite{Robe18}.
Hence, quake events in our cluster model follow the same log-Poisson
process in Eq.\ (1), as directly verified in Ref.~\citep{Boettcher18b}.

Significant particle motion at $\alpha\to\infty$ only occurs when
activated by quake events, so that the distance traveled is proportional
to the integral of the rate in Eq.\ (1). Thus, it predicts cluster
sizes growing logarithmically with time, consistent with experiments
\cite{Yunker09}, and the emergence of dynamic heterogeneity \cite{Becker14}.
Closely related, and experimentally more accessible \cite{Courtland03,Yunker09},
the MSD of particles moving between times $t_{{\rm w}}$ and $t\ge t_{{\rm w}}$
in a dense colloid which is predicted by Eq.\ (1) to grow as $\langle\Delta x^{2}(t,t_{{\rm w}})\rangle\propto\ln(t/t_{{\rm w}})$.
This ``logarithmic diffusion'' was first tested \cite{BoSi09} with
experimental particle tracking data \cite{Courtland03}. Using the
time-lag $\Delta t=t-t_{{\rm w}}$ as variable, one easily obtains
$\langle\Delta x^{2}(\Delta t)\rangle\propto\ln(1+\Delta t/t_{{\rm w}})$
such that the MSD grows linearly as $\sim\Delta t/t_{{\rm w}}\ll1$
and as $\sim-\ln t_{{\rm w}}+\ln\Delta t$ for $\Delta t\gg t_{{\rm w}}$.
This is exactly the behavior we find in cluster model simulations
as well as in more recent tracking experiments by Yunker et al \cite{Robe16}. 

\begin{figure}
\hfill{}\includegraphics[viewport=0bp 0bp 960bp 613bp,clip,width=0.9\columnwidth]{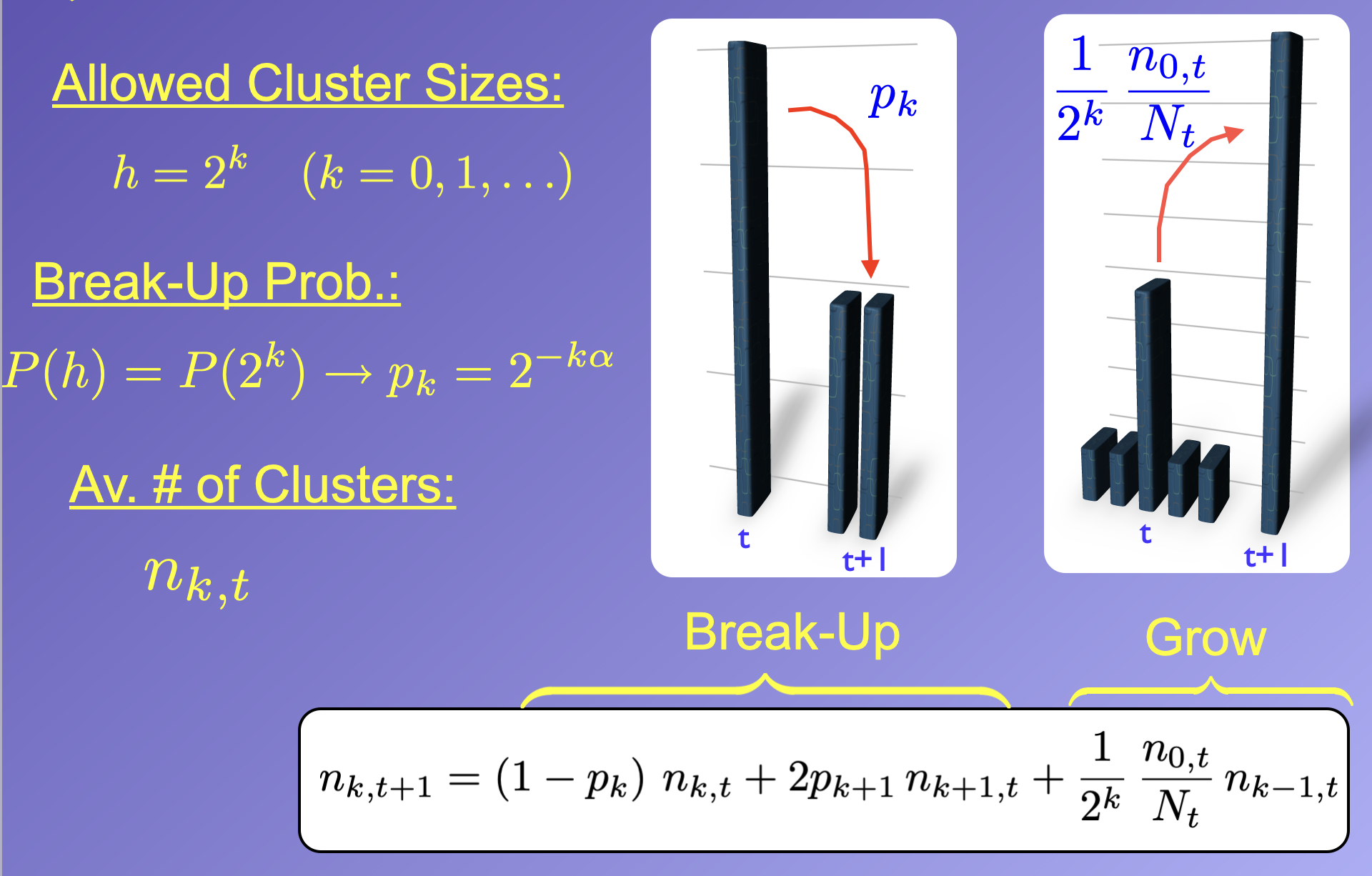}\hfill{}

\caption{\label{fig:clusterrate} Illustration of the construction for the
mean-field rate equation for cluster evolution in Eq. (\ref{eq:DH-1}).}
\end{figure}

\subsection*{Predictions of the Rate Equation}

We have derived a rate equation of the clustering dynamics \cite{BoSi09},
with asymptotic properties that are an apt description of the cluster-size
distribution as a function of time, as shown in Fig.~5b. As illustrated
in Fig. \ref{fig:clusterrate}, the fundamental quantity considered
is the average number of clusters $n_{k,t}$, all abstracted into
strictly \emph{power-of-2} sizes, $h=2^{k}$ ($k=0,1,2,\ldots$),
extant at sweep-time $t$ after the quench at $t=0$. For $k>0$ (i.e.,
$h>1$), those $k$-clusters are immobile and break up with probability
$p_{k}\sim P_{\alpha}(2^{k})$ at $\beta=1$ from Eq.\ (\ref{eq:Ph}),
but strictly into two half-sized ($k-1$)-clusters. To write down
a self-consistent equation for $n_{k,t}$, we stipulate that during
each \emph{sweep} some $k$-clusters $n_{k,t}$ for all $k>0$ are
lost through break-ups with probability $p_{k}$, yet, they also are
gained (with probability $2p_{k+1}$) through the break-up of ($k+1$)-clusters
into \emph{two} $k$-clusters. Unit $k=0$-clusters represent \emph{indivisible
mobile} particles ($h=1$). Larger $k$-clusters grow further via
accretion of those small mobile clusters, but only in groups of $2^{k}$
with probability $1/2^{k}$. During a sweep, \emph{all} $n_{0,t}$-clusters
attach somewhere. We obtain 
\begin{eqnarray}
n_{k,t+1}-n_{k,t} & = & -p_{k}n_{k,t}+2p_{k+1}n_{k+1,t}+\frac{n_{k-1,t}n_{0,t}}{2^{k}N_{t}}\label{eq:DH-1}
\end{eqnarray}
for $k>0$, and $n_{0,t+1}=2p_{1}n_{1,t}$. The last term describes
growth through the accretion of all unit clusters $n_{0,t}$, equally
shared between all clusters, $N_{t}=\sum_{k=0}^{\infty}n_{k,t}$.
Its form ensures particle number conservation, $\sum_{k=0}^{\infty}2^{k}n_{k,t}$,
for all $t$. Indeed, the solution of this equation \cite{BoSi09}
predicts two regimes, with a stationary solution for the ``fluid''
phase when $\alpha\leq1$, while for $\alpha>1$ stationarity is lost
and the cluster-size distribution evolves like a \emph{wave} on a
$t^{1/\alpha}$-clock (i.e., on the $\ln t$-clock of RD for $\alpha\to\infty$),
with an increasing ``gap'' of all clusters of smaller sizes vanishing,
as in Fig.~5c. Asymptotically, we found that average sizes in the
rate equations (and correspondingly the gap) now grow as $\left\langle h\right\rangle \sim t^{1/\alpha}$,
i.e., $\left\langle h\right\rangle \sim\ln t$ for $\alpha\to\infty$,
the physical case. This would indicate a rate of record events growing
as $\lambda(t)\propto\frac{\partial\left\langle h\right\rangle }{\partial t}\sim t^{\frac{1}{\alpha}-1}$,
thus, $\lambda(t)\propto r/t$ for $\alpha\to\infty$. In turn, for
$\alpha=1$, stationarity results (``liquid'').

\section*{Appendix B: Simulations of the Bistable Network Model}

For this work we employed the LAMMPS scripts provided by in the Supplement of Ref. \cite{Shohat23}. Although the authors included one initial condition to run the test cases, the information to reproduce and create new initial conditions, with different bond structures as well as different number of nodes was reconstructed following Ref. \cite{Shohat22}, where they explain the main characteristics of the numerical model and parameters also applied in Ref. \cite{Shohat23}.

According to this reference, the initial conditions were created by simulating a set of soft spheres in a square domain ($x=10-440$, $y =10-440$). Particles of three different sizes ($d=11$,12 and 13 units) were inserted randomly into this domain, and the overlap among them resolved using a FIRE-minimization algorithm (Fast Inertial Relaxation Engine).
Although not explicitly stated in the reference material, we used a wall domain, similarly created by the same distribution of particles in the outer region of this domain, to generate rough walls that enhance the disorder of the system.
Using the mechanically stable configuration of particles obtained after the relaxation, we created configurations of nodes and bonds by detecting the particles in contact. (In the process, the particles are therefore converted to ''nodes'', i.e., the diameter is ignored after this step.)
 
The bond length distribution obtained in this manner follows the one presented in those reference, which are between 9 and 13 distance units. As per the reference,  5 different ''bond-types'' were assigned, each ruled by a different bistable potential (see  bistable\_pot.txt, provided by Ref. \cite{Shohat23},). These are double-well potentials of the form:
\begin{eqnarray}
U &=&\frac{a_4}{4}\delta R^4-\frac{a_2}{2}\delta R^2,
\label{eq:bipotential}
\end{eqnarray}
where $\delta R=R-R^0$, represents the deviation from the mean of the bond's two rest lengths. The parameters $a_4=1$ and $a_2=2.5$ are identical for all bonds, while $R^0\in[9,11]$ is chosen at random for each bond.  

Once the initial configuration is set up, the crumbled sheet simulation proceeds as follows:
bonds on the top and bottom of the system (10 units lengths) are selected as fixed walls that exert constant pressure on the system. These are only allowed to move in the vertical direction. Bonds to the right and left of the domain (10 unit lengths) are only allowed to move horizontally.
The system is left to stabilize under this constant pressure. Due to the
incompatibility between the bonds' rest lengths, the system stabilizes in a local minimum energy configuration, a state that is highly frustrated and carries excess stresses. As a result, many
bonds deviate from their minimal energy state.
After the system reaches this state, we add noise to the simulations in the form of a very small temperature $T$. It is much lower than the value needed to flip any bond between its minima but it is enough as ''to allow a sequences of instabilities in which many bonds shift to their shorter rest length'' \cite{Shohat23}.
Events are defined as these bonds switch between minima, moving across the potential barrier for a distance larger than $dR_{\rm critical} = \sqrt{\frac{a_2}{3a_4}}$.

Although the authors of Ref. \cite{Shohat23} provided the LAMMPS script to run the crumbled sheet simulation, as well as an initial configuration for a system with $N=2000$ nodes (forming 5349 bonds), the script for producing such an initial condition was not part of their supplementary information. Therefore, we had to write our own LAMMPS script  to create the initial configuration of the disordered network by evolving an initial pack of spheres. We used three different system sizes, i.e., number of spheres ($N=2000$, 4000, 8000). We adjusted the area in which these spheres were created, to have a constant packing fraction among different size systems (so that we matched the packing fraction of the initial configuration provided). After the pack of spheres stabilized under the constrains, we created the network of contacts, as explained above. For the subsequent  crumbling sheet simulation, we utilized the LAMMPS script supplied by Ref. \cite{Shohat23}.
We adjusted the force applied by the boundaries for  the different system sizes, scaling it by $\sqrt{l}$, with $l$ being the length of the domain size, in order to achieve the same pressure for systems of different sizes.

The authors of  Ref. \cite{Shohat23} state that ''for the same network topology, many realizations can be generated by 'shuffling' the
randomized lengths at rest, $R^0$'', and that they ''run repeated simulations with the same initial condition (the configuration
at the beginning of the plateau) and the same random selection of bond types, but with different realizations of the thermal fluctuations.''
In our case, we created 10 different network configurations for each
system size in this study, to each of which 100 different configurations of bonds were assigned, resulting in 1000
different systems for each scenario investigated. This should minimize any possible bias given by the system structure.

\bibliographystyle{apsrev4-2}
\bibliography{/Users/sboettc/Boettcher}

\end{document}